# Communication with spatially modulated Light through turbulent Air across Vienna

Mario Krenn[1,2,*], Robert Fickler[1,2], Matthias Fink[2], Johannes Handsteiner[2], Mehul Malik[1,2], Thomas Scheidl[1,2], Rupert Ursin[2], Anton Zeilinger[1,2,*]

[1]Vienna Center for Quantum Science and Technology (VCQ), Faculty of Physics, University of Vienna, Boltzmanngasse 5, A-1090 Vienna, Austria.
[2]Institute for Quantum Optics and Quantum Information (IQOQI), Austrian Academy of Sciences, Boltzmanngasse 3, A-1090 Vienna, Austria.
* mario.krenn@univie.ac.at, anton.zeilinger@univie.ac.at

**The transverse spatial modes of light offer a large state-space with interesting physical properties [1,2]. For exploiting it in future long-distance experiments, spatial modes will have to be transmitted over turbulent free-space links. Numerous recent lab-scale experiments have found significant degradation in the mode quality after transmission through simulated turbulence and consecutive coherent detection [3-7]. Here we experimentally analyze the transmission of one prominent class of spatial modes, the orbital-angular momentum (OAM) modes, through 3 km of strong turbulence over the city of Vienna. Instead of performing a coherent phase-dependent measurement, we employ an incoherent detection scheme which relies on the unambiguous intensity patterns of the different spatial modes. We use a pattern recognition algorithm (an artificial neural network) to identify the characteristic mode pattern displayed on a screen at the receiver. We were able to distinguish between 16 different OAM mode superpositions with only ~1.7% error, and use them to encode and transmit small grey-scale images. Moreover, we found that the relative phase of the superposition modes is not affected by the atmosphere, establishing the feasibility for performing long-distance quantum experiments with the OAM of photons. Our detection method works for other classes of spatial modes with unambiguous intensity patterns as well, and can further be improved by modern techniques of pattern recognition.**

Angular momentum of photons consists of two different components. The first one is the Spin Angular Momentum (SAM), which defines the polarization of photons. The second component is the Orbital Angular Momentum (OAM), which corresponds to the spatial phase distribution of the photon. Both components have been used extensively in optical experiments at the lab-scale. Furthermore, polarization has been successfully used in quantum experiments over free-space links on the order of 100 kilometers [8-10]. The polarization of a photon, while being easily controllable and immune against atmospherical influences, resides in a two-dimensional state-space. This places an inherent limit on how much information one can send per photon. As one consequence, it sets a tight bound on how much error a quantum key distribution (QKD) system that uses such encoding can tolerate [11,12]. An alternate way to encode information is in the OAM degree-of-freedom of a photon, which offers a theoretically unbounded number of discrete levels [1,2], and is able to improve classical [13-15] as well as quantum communication [16,17]. Light carrying OAM has a "twisted" or helical wave front, with an azimuthal phase varying from 0 to $2\pi\ell$. The integer $\ell$ stands for the topological charge or helicity, and $\ell\hbar$ is the OAM of the photon [18].

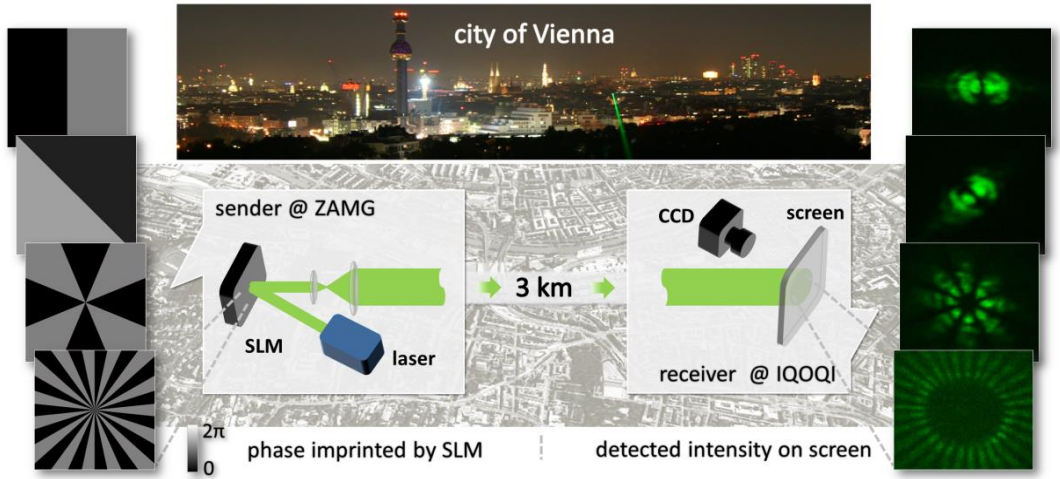

**Figure 1**: Sketch of the experimental setup. The 3 kilometer free-space experiment was performed in the city of Vienna, from ZAMG (*Zentralanstalt für Meteorologie und Geodynamik*, Central Institute for Meteorology and Geodynamics) to our institute IQOQI. Top: Picture of an alignment laser from IQOQI to ZAMG, captured at ZAMG. Left: The sender modulates a 532nm laser with an SLM (Spatial Light Modulator). Different phase holograms that modulate the beam are shown; they correspond to superpositions of OAM modes (from top to bottom) with $\ell=\pm 1$, $\ell=\pm 1$ rotated, $\ell=\pm 4$ and $\ell=\pm 15$. Right: At the receiver we observe the transmitted modes and record them with a CCD camera. The images correspond to the modulated phases on the left. The size of the inner most intensity ring varies from approximately 12cm for $|LG_{\pm 1}\rangle$ to 72cm for $|LG_{\pm 15}\rangle$. By analyzing the observed images, we characterize the atmospheric stability of the OAM modes and use them for transmitting real information. (Geographic pictures taken from Google Earth, ©2014 Google, Cnes/Spot Image, DigitalGlobe)

The ability to transmit OAM modes over free-space is crucial for taking such lab-scale experiments into the real world. Several theoretical studies have analyzed the behavior of OAM light beams in turbulent atmosphere [19-24]. Based on these studies, atmospheric turbulence has been simulated in the laboratory using spatial light modulators, heat pipes, and rotating phase plates [3-5]. The results give reason to assume that long-distance free-space transmission of OAM modes is very challenging or even unfeasible, since refractive index fluctuations lead to severe crosstalk at the receiver. Recently, this detrimental effect of atmosphere has been confirmed by two laboratory experiments simulating a 1km turbulent path [6,7]. Whereas experiments with twisted radio waves for classical data transmission have been performed over ~450m in a free-space link in a city [25], the same has not yet been possible for optical frequencies because of the sensibility of visible light to turbulence. The only two experiments that go beyond typical laboratory distances are [26], which analyses classical OAM mode transmission over 15meters, and [27] which uses a polarization-OAM hybrid system for quantum key distribution over 210meters. The latter experiment has been performed in a big hall, to minimize the difficulties with turbulent atmosphere.

The studies mentioned above employed coherent mode detection. Different OAM beams are generated from Gaussian beams using holographic transformations at the sender and then transmitted through simulated atmosphere. At the receiver, the transmitted modes are back transformed to Gaussian beams in order to analyze the quality of the OAM modes after free-space propagation. However, such transformations [28-30] rely on axis-dependent decomposition of modes and are extremely sensitive to angle of arriving (AOA) fluctuations and beam wander. As the effect of atmospherical turbulences on the beam propagation can be decomposed into different orders of Zernike-polynomials [31], the difficulty in using such techniques can be explained rather intuitively. Beam wander (lateral tilts) is the first order contribution of the atmosphere, which influences such phase measurement based detection schemes significantly. Additionally, higher order contributions such as defocusing, astigmatism or coma further reduces the mode transformation quality and efficiency, for instance due to inefficient identification of a defocused Gaussian mode.

In this work we transmitted OAM superposition modes of light over a 3km intra-city link in Vienna under strong-turbulence conditions. We employ an incoherent detection scheme by directly observed the unambiguous mode-intensity patterns on a screen with the help of a standard adaptive pattern recognition algorithm. This method avoids coherent phase-dependent measurements for identifying the transmitted OAM modes, therefore is not affected by the atmospherical contribution described above. We transmitted 16 different mode superpositions ($l=\pm 0, \pm 1, \ldots \pm 15$) and could distinguish them with an average error of only ~1,7%. Additionally, we analyzed the atmospheric effect on the relative phase of these superpositions, which is a crucial property for verifying quantum entanglement of OAM modes in future experiments. We found that the relative phase is only slightly affected by turbulence which favors the use of petal patterns in the experiment. The experiment exploiting the stability of relative phases to verify long-distance OAM entanglement could use a polarization-OAM hybrid entangled state: The polarization is measured locally, and triggers a distant ICCD which detects the OAM photon, similar to the analogue lab-scale experiment in [32].

We have chosen OAM superposition modes due to their interesting physical properties and well developed methods to detect quantum entanglement in future experiments. However, this approach does not only work for OAM superposition modes, but for every spatial mode structures with unambiguous distinguishable intensity patterns. The highly symmetric petal patterns of OAM superpositions seem to be advantageous in turbulent conditions in order to correctly identify the modes with a pattern recognition algorithm. It might be interesting to analyze whether different spatial mode structures (such as Hermite-Gaussian beams, self-healing Bessel beams [33] or generalizations such as Ince-Gaussian modes [34]) are identifiable in a more robust way. Another interesting open question is the scaling of the pattern recognition quality with the number of photons detected, namely how many photons are required to identify an intensity pattern within a given error, and how turbulence affects these numbers. It will be interesting to investigate this question in a future experiment.

The experimental setup for sending and receiving OAM modes over the city of Vienna can be seen in Figure 1. The sender was located in a ~35m high radar tower of *Zentralanstalt für Meteorologie und Geodynamik*, (ZAMG, Central Institute for Meteorology and Geodynamics). The receiver was on the rooftop of our institute building 3040 meters away.

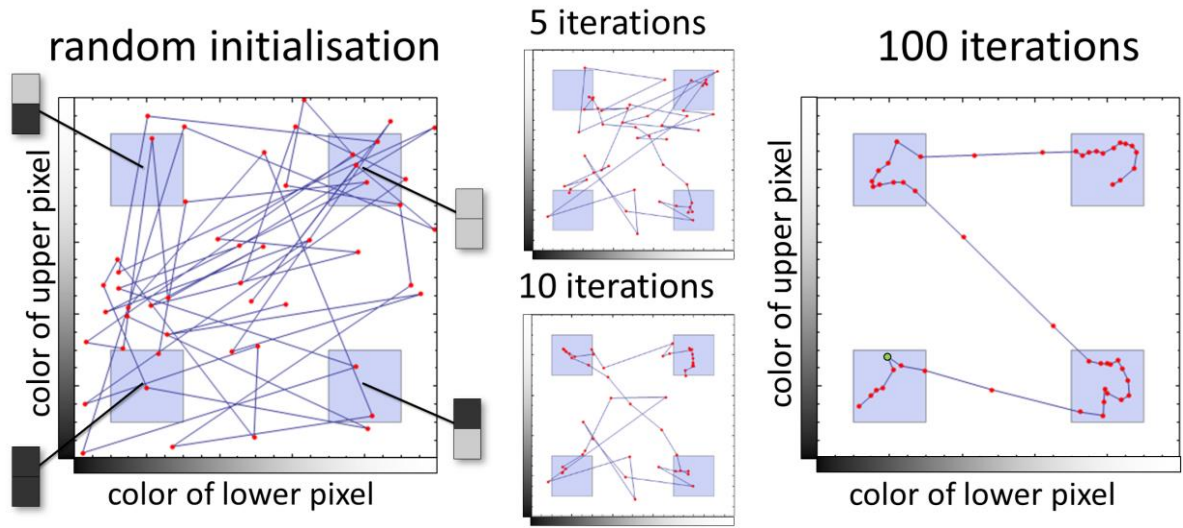

**Figure 2**: Working principle of the artificial neuronal network (ANN) described by an example of four different grey-scale images with only two pixels each. A network with a chain of 50 neurons is initialized (left). The neurons (red dots) are connected to their neighbors, as indicated by the blue lines. Every neuron is assigned a random position in 2-dimensional space which corresponds to their grey-value (the color of the upper pixel corresponds to a position at the y-dimension; the lower one corresponds to the x-axis). In each subsequent step, the ANN receives one out of four possible 2-pixel images (dark-dark, dark-bright, bright-dark, bright-bright), which as well corresponds to a coordinate in the 2-dimensional space. Due to "virtual turbulence", the color of the images might be changed slightly, resulting in a 2-dimensional region (indicated in blue) rather than a single position. For each input image, the position of the *winner neuron* (the neuron closest to the input image) is pulled strongly towards the input image. The positions of the *winner neuron*'s neighbors are pulled towards the input picture as well – but less significantly. After a few iterations (middle), the initially random network organizes itself and finds the structure of the input in an unsupervised way (right). After the training phase, we can assign to each neuron the information to which image it corresponds (for example, the green neuron in the right picture belongs to the image dark-dark). For this we input perturbed images (for instance a dark-dark image), and assign the information about the image to the *winner neuron* (for example, to the green neuron). As soon as most neurons have been assigned a specific image information, the network can be used for identifying unknown images. To identify unknown images, we again calculate the *winner neuron* for each image, and get the image information assigned to the neuron. In the example above, 5 out of 50 neurons in the final network lie outside the blue area, and it is likely that they will never be a *winner neuron*, thus could be removed from the network (for computational speedup). In our experiment, the inputs are images with 720x720 = 518.400pixels, thus the size of the virtual neuronal network space is 518.400-dimensional. In that space, the ANN autonomously categorizes the 16 input structures that describe our different OAM modes-superpositions.

At the sender, we had a 20mW laser with a wavelength of 532nm. The light from the laser was modulated by a Spatial Light Modulator (SLM) which impressed the phase information of the OAM modes onto the beam. Then it was expanded by a telescope to a diameter of approximately 6cm and sent with a high-quality f=30cm lens to the receiver. The received mode intensity profiles were observed on a screen, and recorded with a CCD camera.

We characterize the turbulence of the atmosphere in the following way: The atmospheric turbulence can be decomposed in cells with similar pressure, thus similar refractive index. Those "Fried cells" are flowing across the propagating path and randomly deflect the centroid of the beam in different directions. If the light beam is observed on a screen it "jumps" around on timescales of some 1 kHz. However, when averaged over time, the intensity's centroid position remains small. The magnitude of the short-term beam wander depends on the refractive index structure parameter and the path length L. It is characterized by the root-mean-square of the beam displacement from the time-averaged center. We measure the root-mean-square of Gaussian beam size with images of 1/2000sec and 1/4000sec integration time, and with images of 20 seconds integration time. From these values we calculated a Fried parameter of $r_0$=1.4cm and an atmospheric structure constant of $C_n^2$=7.7*10$^{-15}$ m$^{-2/3}$ using the Kolmogorov theory of atmospheric turbulence [35-37]. These parameters correspond to strong turbulence conditions [6,7].

Mathematically, OAM modes of light can be described by Laguerre-Gauss functions $|LG_\ell\rangle$ with a spiral phase distribution exp($i\ell\varphi$), where $\ell$ defines the OAM mode number of photons which can take any integer value [13]. In our experiment we sent light in superpositions of higher-order OAM modes. The complete state can be denoted by

$$|LG^\alpha_{\pm\ell}\rangle = \frac{1}{\sqrt{2}}(|LG_{+\ell}\rangle + e^{i\alpha}|LG_{-\ell}\rangle), \qquad (1)$$

where $\alpha$ denotes the relative phase between the two modes, which corresponds to a rotation of the phase and intensity structure. The transverse phase structure of an OAM-superposition $|LG_{\pm\ell}\rangle$ is radially symmetric and has $2\ell$ phase jumps of $\pi$ in the azimuthal direction. Its intensity distribution shows $2\ell$ maxima and minima arranged in a ring (Figure 1).

The size of the detected modes were between 12cm for $|LG_{\pm1}\rangle$ to 72cm for $|LG_{\pm15}\rangle$, were the final beam size variation is mainly due to diffraction. The size of the intensity of our modes scales linearly with $\ell$, as it is expected for holographic generation without intensity shaping [38]. The size can be reduced significantly by applying intensity shaping on the SLM or using non-diffractive beams such as Bessel beams. In our experiment however, beam size was not an issue.

To analyze the intensity structure of the received modes, we use a standard adaptive pattern recognition algorithm in form of an artificial neural network [39,40]. In the "training" or initialization phase, the algorithm receives a number of recorded images in order to autonomously learn how to recognize the different patterns. After this initialization, the algorithm is ready to analyze the real data in form of images. The working principle is explained with a simplified example in Figure 2. As the neural network is trained with images involving atmosphere-induced disturbances, it is developing automatically a robust detection despite such effects. A significant advantage of our method is that no further image corrections need to be performed.

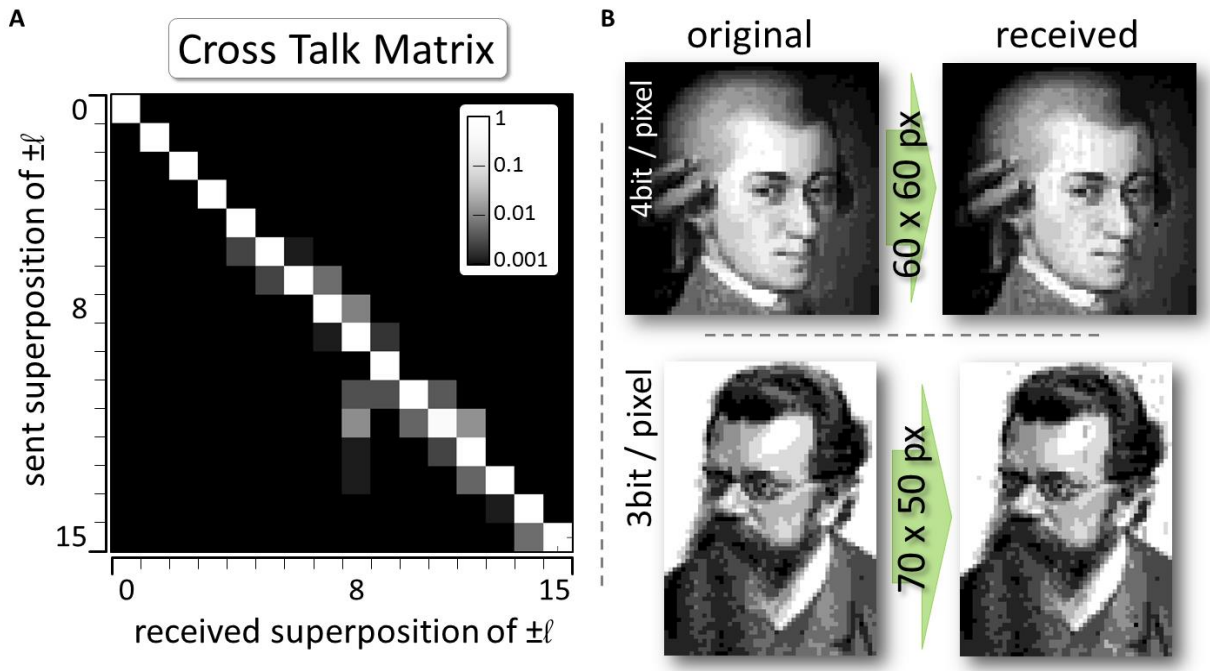

**Figure 3**: A) The crosstalk matrix for different OAM superposition settings (in logarithmic scale). The cross talk matrix shows the distinguishability of OAM mode superpositions from $|LG_0\rangle$ to $|LG_{\pm 15}\rangle$ (corresponding images in figure 1). We can distinguish the modes with an average error rate of 1.7% and a channel capacity of 3.89bits. B) We transmit two greyscale images encoded in these OAM-mode superpositions. The upper image (Wolfgang Amadeus Mozart) has 4 bits per pixel, which corresponds to 16 greyscale settings. As a result, the full available set of modes was used to encode this image. The received image has a bit-error-ratio of 1.2%. The lower image (Ludwig Boltzmann) has 3 bits per pixel, which required 8 different modes. The average error rate for this image is measured to be 0.8%.

In the first step, we investigate whether the characteristic mode patterns of different modes can be distinguished after free-space transmission. For that we analyze the crosstalk between the first 16 OAM mode-superpositions $|LG_{\pm \ell}\rangle$ (with $\alpha=0$), from $\ell=0$ up to $\ell=15$, at the receiver. For each transmitted $\ell$-value, we accumulated approximately 450 received mode-intensity images, which served as the input for our algorithm. As a result, we obtained the detected $\ell$-value of the received mode. By comparing the prepared and measured $\ell$-values, we can calculate the crosstalk matrix between the different OAM modes, which is shown in Figure 3A. For superpositions of $|LG_0\rangle$ up to $|LG_{\pm 15}\rangle$ we find a good distinguishability with an average error of 1.7%. The error is defined as the ratio between wrong detected modes and all detected modes.

To illustrate the quality of the received modes, we used these 16 different states for encoding two grayscale images with 8 and 16 different greyscale values (corresponding to 3bits and 4bits per pixel, respectively). Each transmitted mode carried the information of one pixel of the image. The mode $|LG_0\rangle$ corresponds to black, while higher-order modes correspond to grey-values. The highest mode in the alphabet ($|LG_{\pm 7}\rangle$ for 3bits and $|LG_{\pm 15}\rangle$ for 4bits) corresponds to white. The maximum frame rate of the SLM is 60Hz, and for the camera we used 50Hz. As the SLM and the CCD camera were not synchronized, and the liquid crystal display of the SLM was slow due to the low environment temperature of ~5-10°C, we displayed each mode for 10 SLM frames. In between two modes, the beam was deflected for 4 frames to distinguish between the subsequent modes. The transmission rate thus was 4 pixel/sec. The transmitted and decoded image can be seen in figure 3B. The errors in the decoded 3bit and 4bit images are 0.8% and 1.2%, respectively. The bit-error is defined as the ratio between wrong bits in the decoded image and all bits of the images.

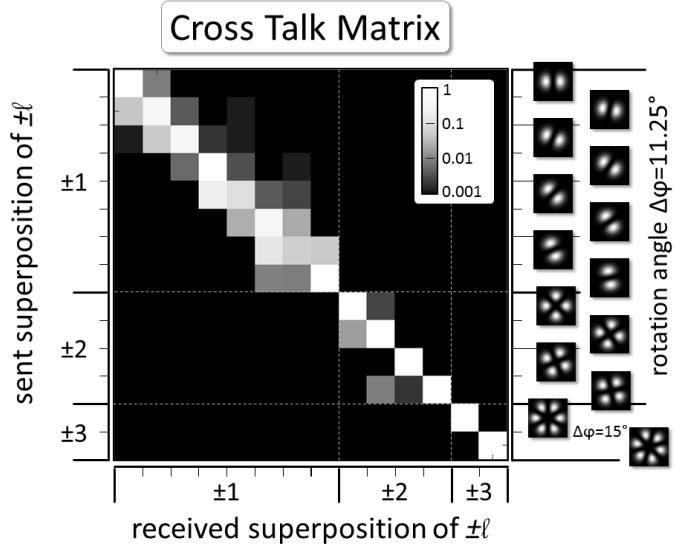

**Figure 4**: Cross Talk Matrix for rotated mode patterns (in logarithmic scale). We send OAM superpositions of |LG$_{±1}$⟩, |LG$_{±2}$⟩, and |LG$_{±3}$⟩ with different relative phases, which leads to rotated intensity patterns by $\Delta\varphi$. The pattern recognition algorithm is able to distinguish them with an error-ratio of 0.7% for angles of 22.5° and an error of 15.9% for angles as small as 11.25°. The stability of relative phases is crucial for follow-up experiments on quantum entanglement of OAM over large distances.

As mentioned above, the relative phase $\alpha$ of OAM superpositions is crucial for verifying quantum entanglement of OAM modes in potential future experiments. Hence, we analyzed in a second step the stability of $\alpha$ under atmospheric turbulence. By changing the relative phase, one can access the whole equator of the qubit Bloch sphere, thus has access to two different mutually unbiased bases (MUBs). A high visibility in those two MUBs would be sufficient to verify entanglement [41]. For OAM modes, the relative phase rotates the intensity pattern of the mode, which we can directly observe.

We changed the relative phase in modes |LG$^\alpha_{±1}$⟩, |LG$^\alpha_{±2}$⟩ with step sizes of 11.25°, and analyzed their distinguishability with the pattern recognition algorithm (Figure 4). With this angle, the rotated modes could be distinguished with an error of 15.9% averaged for |LG$^\alpha_{±1}$⟩ and |LG$^\alpha_{±2}$⟩. When we omit every second measured phase angle, and analyze the relative phases in 22.5° steps, we find an average error of only 0.7%.

The smaller error in the case of a rotation angle 22.5° than for 11.25° is expected because the intensity patterns are rotated by twice the angle. Therefore they are easier to distinguish. The small influence of the atmosphere on the relative phase α indicates the feasibility of quantum entanglement experiments with OAM over long distances. For such an experiment, we would need to extent our method to observe coincidence intensity patterns at the single-photon level [42]. We expect this to be possible by replacing our screen with a triggered ICCD camera, without any mode transformation, as shown in [32]. Whereas it is not known whether this detection method might be useful for quantum key distribution, it has the potential to detect interesting fundamental properties such as quantum entanglement.

In summary, we showed that information encoded in the OAM property of light can be extracted after propagation through a 3 kilometre intra-city link with strong turbulences. We characterized the rotational mode stability and found it to be very resistant to degradation by the atmosphere. Furthermore, it was possible to distinguish 16 different OAM modes, which we used to encode small grey-scale images. Our findings also indicate that possible future long-distance quantum experiments with the OAM degree of freedom will be possible, even without adaptive optical systems [42, 32]. Furthermore, as the atmosphere does not destroy the modal structure, it should be sufficient to use tip-tilt correction at the receiving telescope in order to use conventional holographic OAM-detection methods [28-30]. The horizontal distance of 3km of our transmission link is on the order of the effective vertical thickness of the atmosphere of 6km [43], which suggests that earth-to-satellite links using OAM modes are not limited by atmospherically turbulence.

**Acknowlegdement:**
We thank an anonymous reviewer for interesting suggestions to our manuscript, Roland Potzmann and ZAMG for providing access to the radar tower, and Adam Wyrzykowski and William Plick for help with the experiment. This work was supported by the European Commission (SIQS, No. 600645 EU-FP7-ICT); and the Austrian Science Fund FWF with SFB F40 (FoQus). MM acknowledges support form the European Commission through a Marie Curie Fellowship.

**References**
[1] Torres, J. P. & Torner, L., Twisted Photons: Applications of Light with Orbital Angular Momentum. Bristol: Wiley-VCH (2011).

[2] Andrews, D. L. & Babiker, M., The Angular Momentum of Light. Cambridge: Cambridge University Press. (2012).

[3] Pors, B. J., Monken, C. H., Eliel, E. R., & Woerdman, J. P., Transport of orbital-angular-momentum entanglement through a turbulent atmosphere. *Optics Express*, **19**(7), 6671-6683 (2011).

[4] Malik, M., *et al.*, Influence of atmospheric turbulence on optical communications using orbital angular momentum for encoding. *Optics Express* **20**(12), 13195-13200 (2012).

[5] Ibrahim, A. H., Roux, F. S., McLaren, M., Konrad, T., & Forbes, A., Orbital-angular-momentum entanglement in turbulence. *Physical Review A*, **88**(1), 012312 (2013).

[6] Rodenburg, B., *et al.*, Simulating thick atmospheric turbulence in the lab with application to orbital angular momentum communication. *New J. Phys,.* **16**, 033020 (2014).

[7] Ren, Y., *et al.*, Atmospheric turbulence effects on the performance of a free space optical link employing orbital angular momentum multiplexing. *Optics letters*, **38**(20), 4062-4065 (2013).

[8] Ursin, R., *et al.*, Entanglement-based quantum communication over 144 km. *Nature Physics*, **3**(7), 481-486 (2007).

[9] Yin, J., et al., Quantum teleportation and entanglement distribution over 100-kilometre free-space channels. *Nature*, **488**(7410), 185-188 (2012).

[10] Ma, X. S., et al. Quantum teleportation over 143 kilometres using active feed-forward. Nature, **489**(7415), 269-273 (2012).

[11] Bourennane, M., Karlsson, A., Bjork, G., Gisin, N., & Cerf, N., Quantum key distribution using multilevel encoding: security analysis. Journal of Physics a-Mathematical and General, 35(47), 10065–10076 (2002).

[12] Huber, M., & Pawłowski, M., Weak randomness in device-independent quantum key distribution and the advantage of using high-dimensional entanglement. Physical Review A, 88(3), 032309 (2013).

[13] Wang, J., et al., Terabit free-space data transmission employing orbital angular momentum multiplexing, *Nature Photonics* **6**, 488–496 (2012).

[14] Bozinovic, N., et al., Terabit-scale orbital angular momentum mode division multiplexing in fibers. *Science*, **340**(6140), 1545-1548 (2013).

[15] Xie, G., Ren, Y., Huang, H., Lavery, M. P., Ahmed, N., Yan, Y., ... & Willner, A. Experiment Turbulence Compensation of 50-Gbaud/s Orbital-Angular-Momentum QPSK Signals using Intensity-only based SPGD Algorithm. In Optical Fiber Communication Conference (pp. W1H-1). Optical Society of America (2014).

[16] Gröblacher, S., Jennewein, T., Vaziri, A., Weihs, G., & Zeilinger, A., Experimental quantum cryptography with qutrits. *New Journal of Physics*, **8**(5), 75 (2006).

[17] Mirhosseini, M., Magaña-Loaiza, O. S., O'Sullivan, M. N., Rodenburg, B., Malik, M., Gauthier, D. J., & Boyd, R. W. High-dimensional quantum cryptography with twisted light. arXiv:1402.7113 (2014).

[18] Allen, L., Beijersbergen, M. W., Spreeuw, R. J. C., & Woerdman, J. P., Orbital angular momentum of light and the transformation of Laguerre-Gaussian laser modes. *Physical Review A*, **45**(11), 8185-8189 (1992).

[19] Paterson, C., Atmospheric turbulence and orbital angular momentum of single photons for optical communication. *Physical review letters*, **94**(15), 153901 (2005).

[20] Anguita, J. A., Neifeld, M. A., & Vasic, B. V., Turbulence-induced channel crosstalk in an orbital angular momentum-multiplexed free-space optical link. *Applied optics*, **47**(13), 2414-2429 (2008).

[21] Tyler, G. A., & Boyd, R. W., Influence of atmospheric turbulence on the propagation of quantum states of light carrying orbital angular momentum. *Optics letters*, **34**(2), 142-144 (2009).

[22] Chandrasekaran, N. & Shapiro, J. H., Photon information efficient communication through atmospheric turbulence—Part I: Channel model and propagation statistics. *J. Lightw. Technol.*, **32**(6), 1075–1087 (2014).

[23] Chandrasekaran, N., Shapiro, J. H., & Wang, L., Photon Information Efficient Communication Through Atmospheric Turbulence—Part II: Bounds on Ergodic Classical and Private Capacities. *J. Lightw. Technol.*, **32**(6) 1088–1097 (2014).

[24] Shapiro, J. H., Comment on `Simulating thick atmospheric turbulence in the lab with application to orbital angular momentum communication', *arXiv:1406.2557* (2014).

[25] Tamburini, F., Mari, E., Sponselli, A., Thidé, B., Bianchini, A., & Romanato, F. (2012). Encoding many channels on the same frequency through radio vorticity: first experimental test. *New Journal of Physics,* **14**(3), 033001.

[26] Gibson, G., et al., Free-space information transfer using light beams carrying orbital angular momentum, Optics Express, 12(22), 5448-5456 (2004).

[27] Vallone, G., D'Ambrosio, V., Sponselli, A., Slussarenko, S., Marrucci, L., Sciarrino, F., and Villoresi, P., Free-Space Quantum Key Distribution by Rotation-Invariant Twisted Photons, *Phys. Rev. Lett.* **113**, 060503 (2014).

[28] Mair, A., Vaziri, A., Weihs, G., & Zeilinger, A., Entanglement of the orbital angular momentum states of photons. *Nature*, **412**(6844), 313-316 (2001).

[29] Berkhout, G. C., Lavery, M. P., Courtial, J., Beijersbergen, M. W., & Padgett, M. J., Efficient sorting of orbital angular momentum states of light. *Physical review letters*, **105**(15), 153601 (2010).

[30] Mirhosseini, M., Malik, M., Shi, Z., & Boyd, R. W., Efficient separation of the orbital angular momentum eigenstates of light. *Nature communications*, **4** (2013).

[31] Noll, R. J., Zernike polynomials and atmospheric turbulence, *JOSA* **66**(3), 207-211 (1976).

[32] Fickler, R., Krenn, M., Lapkiewicz, R., Ramelow, S., & Zeilinger, A., Real-Time Imaging of Quantum Entanglement. *Scientific reports*, **3** (2013).

[33] McLaren, M., Mhlanga, T., Padgett, M. J., Roux, F. S., & Forbes, A. Self-healing of quantum entanglement after an obstruction. *Nature communications*, **5** (2014).

[34] Bandres, M. A., JC Gutiérrez-Vega, J. C., Ince gaussian beams, Optics letters 29 (2), 144-146 (2004)

[35] Kolmogorov, A. N., The local structure of turbulence in incompressible viscous fluid for very large Reynolds numbers. *Dokl. Akad. Nauk SSSR*, **30**(4), 299-303 (1941).

[36] Fante, R. L., Electromagnetic beam propagation in turbulent media: an update. *Proceedings of the IEEE*, **68**(11), 1424-1443 (1980).

[37] Fried, D. L., Optical resolution through a randomly inhomogeneously medium. *JOSA* **56**, 1372 (1966).

[38] Curtis, J. E., Grier, D. G., Structure of Optical Vortices, *Phys. Rev. Lett.* **90**, 133901 (2003).

[39] Kohonen, T. Self-organized formation of topologically correct feature maps. *Biological cybernetics*, **43**(1), 59-69 (1982).

[40] Kohonen, T. Self-organizing maps (Vol. 30). Springer (2001).

[41] Gühne, O., Toth, G., Entanglement detection, *Physics Reports* **474**, 1, (2009).

[42] Fickler, R., *et.al*, Quantum entanglement of high angular momenta. *Science*, **338**(6107), 640-643 (2012).

[43] Bohren, C. F. & Albrecht, B. A., Atmospheric Thermodynamics, Oxford University Press, New York, 1988.